\documentstyle[twocolumn,prl,aps,epsfig,psfig]{revtex}
\begin{document}
\twocolumn[\hsize\textwidth\columnwidth\hsize\csname@twocolumnfalse%
\endcsname
\title{Non-Perturbative Aspects of Fano Resonances in Quantum Dots: An Exact Treatment}
\author{Robert M. Konik}
\address{Department of Physics, University of Virginia, Charlottesville,
VA 22903}

\date{\today}
\maketitle
\begin{abstract}
We consider transport through quantum dots with two tunneling paths.
Interference between paths gives rise to Fano resonances exhibiting 
Kondo-like physics.  In studying such quantum
dots, we employ a generalized Anderson model which we argue to be
integrable.  The exact solution is non-perturbative in the 
tunneling strengths of both paths.
By exploiting this integrability, we compute the zero temperature
linear response conductance of the dot and so obtain 
reasonable quantitative agreement with the experimental measurements
reported in G\"ores et al. PRB 62, 2188 (2000).  
\end{abstract}
\pacs{PACS numbers: ???}  ]  
\bigskip
\vskip .4in

\newcommand{\del}{\partial}
\newcommand{\ep}{\epsilon}
\newcommand{\clsd}{c_{l\sig}^\dagger}
\newcommand{\cls}{c_{l\sig}}
\newcommand{\cesd}{c_{e\sig}^\dagger}
\newcommand{\ces}{c_{e\sig}}
\newcommand{\up}{\uparrow}
\newcommand{\down}{\downarrow}
\newcommand{\il}{\int^{\tilde{Q}}_Q d\la~}
\newcommand{\ilp}{\int^{\tilde{Q}}_Q d\la '}
\newcommand{\ik}{\int^{B}_{-D} dk~}
\newcommand{\ila}{\int d\la~}
\newcommand{\ilpa}{\int d\la '}
\newcommand{\ika}{\int dk~}
\newcommand{\tQ}{\tilde{Q}}
\newcommand{\rh}{\rho_{\rm bulk}}
\newcommand{\ri}{\rho_{\rm imp}}
\newcommand{\sh}{\sig_{\rm bulk}}
\newcommand{\si}{\sig_{\rm imp}}
\newcommand{\rph}{\rho_{p/h}}
\newcommand{\sph}{\sig_{p/h}}
\newcommand{\rp}{\rho_{p}}
\newcommand{\sip}{\sig_{p}}
\newcommand{\drph}{\delta\rho_{p/h}}
\newcommand{\dsph}{\delta\sig_{p/h}}
\newcommand{\drp}{\delta\rho_{p}}
\newcommand{\dsp}{\delta\sig_{p}}
\newcommand{\drh}{\delta\rho_{h}}
\newcommand{\dsh}{\delta\sig_{h}}
\newcommand{\enp}{\ep^+}
\newcommand{\enm}{\ep^-}
\newcommand{\enpm}{\ep^\pm}
\newcommand{\enph}{\ep^+_{\rm bulk}}
\newcommand{\enmh}{\ep^-_{\rm bulk}}
\newcommand{\enpi}{\ep^+_{\rm imp}}
\newcommand{\enmi}{\ep^-_{\rm imp}}
\newcommand{\enh}{\ep_{\rm bulk}}
\newcommand{\eni}{\ep_{\rm imp}}
\newcommand{\sig}{\sigma}
\newcommand{\la}{\lambda}

In recent years advances in nanotechnology have made possible the fabrication
of single electron transistors (SETs).  SETs, or colloquially, quantum dots,
are characterized by the remarkable ability to tune, via a gate voltage,
the number of localized electrons sitting upon the dot.  By
tuning the number of electrons to be nearly one, a novel Kondo
system is created.  This new realization of an old physical paradigm
has sparked tremendous experimental interest, i.e. \cite{gold,kondo}.

In the simplest realization of quantum dots there exists a single
conduction path through the dot.  However the flexibility inherent
in semi-conductor SETs allows for more exotic scenarios.  Dots
may be fabricated such that multiple tunneling paths are activated.
Generically the presence of multiple tunneling paths leads to
interference effects observable in transport properties.  In the presence
of two tunneling paths, one resonant (i.e. energy dependent),
one not, Fano resonances can arise \cite{fano}.  Such resonances
in quantum dots have been observed \cite{gores,zach} in the form of asymmetric
peaks and dips in the linear response conductance as a function of
the gate voltage.

Although this letter will focus upon the observations of \cite{gores},
Fano resonances are ubiquitous in nanodevices.  Quantum dots have been 
embedded in multiple connected geometries permitting precise delineation of
possible tunneling paths.  Such geometries, in addition, allow threading
by Aharonov-Bohm fluxes.  The behaviour of transport in such devices
has been studied in \cite{kobayashi}.
Fano resonances have also been observed in STM measurements of adsorbed 
magnetic atoms on metallic substrates.  Here interference occurs
because of an interplay between the Kondo resonance and tunneling 
into the continuum of surface conduction electrons \cite{ujsaghy,madhavan}.

The appearance of Fano resonances in quantum dots occurs in conjunction with
Kondo-like phenomena.  Fano resonances
reported in \cite{gores} shows a logarithmic dependence upon
temperature reminiscent of the Kondo effect.  In addition
the authors of \cite{gores} observe a sharp dependence of the
Fano resonances upon small magnetic fields.  Although attributed to
a loss of coherent transport in the resonant scattering channel, 
it might also represent the destruction of a putative Kondo
effect in the dots.  Observations of Fano resonances in STM
tunneling experiments \cite{madhavan} are directly related 
to a Kondo resonance arising from the proximity of
a magnetic adatom.

To model Fano resonances we generalize the 
standard two lead Anderson model in the simplest possible way by 
adding a direct lead-lead coupling.  In the standard 
model electrons
transit from one lead to the other through hopping on and off the dot.
The direct lead-lead tunneling provides a competing scattering path,
nominally independent of energy.  The model Hamiltonian takes the form
\begin{eqnarray}\label{ei}
{\cal H} &=& H_{\rm Anderson} + H_{\rm lead-lead~tunneling}\cr
&=& -i \sum_{l=L,R;~\sig =\uparrow ,\downarrow} 
\int^\infty_{-\infty} dx \clsd (x) \del_x \cls (x) \cr
&& + V_{dl}( \clsd (0) d_\sig + {\rm h.c.})
+ \ep_d \sum_\sig n_\sig + U n_\up n_\down \cr
&& + V_{LR}(c^\dagger_{L\sigma}(0)c_{R\sigma}(0) + {\rm h.c.}).
\end{eqnarray}
Here ${\cal H}$ encodes the standard Anderson model together with
an additional term allowing electrons to transit directly from
one lead to the other.  The $c_{l\sigma}$'s/$d_{\sigma}$'s specify
the lead/dot electrons with $n_\sigma = d^\dagger_\sigma d_\sigma$.
$U$ measures the Coulomb repulsion on the dot while $\epsilon_d$
gives the dot single particle energy.  $V_{ld}$ are the dot-lead hopping
matrix elements.  Experimental realizations of quantum dots generically
see $V_{Ld}\neq V_{Rd}$.  $V_{LR}$ marks the strength of the
direct transmission channel.  The spatial variable $x$ runs from
$-\infty$ to $\infty$ reflecting the `unfolding' of the leads
\cite{long}.

In employing this model to describe the transport properties of
dots such as those studied in \cite{gores},
we assume that only a single level on the dot is relevant to
transport.  This requires the level broadening, $\Gamma \sim V^2$,
to be considerably less than $U+\Delta \epsilon$, where 
$\Delta \epsilon$ is the level spacing.  At least for a subset
of the data in \cite{gores}, this condition is met
with $\Gamma/(U+\Delta \epsilon) \sim .1$.  We note that in
experimental measurements on dots with a single tunneling
path\cite{gold}, $\Gamma/(U+\Delta \epsilon) \sim 1/6$, yet
the Anderson model does an excellent job of describing the
scaling behaviour of the reported finite temperature linear
response conductance.

The model also presumes that the second tunneling path is simply
due to lead-lead hopping.  While it is not entirely clear the 
quantum dots of \cite{gores} are so described, we will show
this model more than adequately describes the observed phenomena.
For dots embedded in a multiply connected geometry,
i.e. \cite{kobayashi}, the ambiguity is lifted and this term provides
a precise description of the second tunneling path.  Fano resonances
in quantum dots were described using random matrix theory in
\cite{clerk} where the exact nature of the direct path need
not be specified.  Aspects of the observations in Ref. \cite{gores}
were described by this treatment.  However Coulomb interactions
were unable to be dealt with directly.  As we will discuss we
believe an exact treatment of the non-perturbative physics present
at finite $U$ is necessary to describe the observations.

The two lead Anderson model with lead-lead tunneling has been studied
previously \cite{bulka,hofstetter}.  
There the model was analyzed by expressing all
the relevant correlators in terms of the dot Greens function 
$\langle d^\dagger d \rangle$ via a system of Dyson equations.  
The dot correlators are then
computed via an equations of motion technique \cite{bulka}, or
a numerical renormalization group \cite{hofstetter}.  
In our exact treatment of the
model we find qualitative differences with this approach.
We believe this is a result of the non-perturbative
physics inherent in the problem.  Although the Dyson equations sum up
all diagrams, they assume nonetheless the problem to be perturbative
in the lead-lead coupling, $V_{LR}$.  Our Bethe ansatz solution indicates
this to not be the case.  In particular for $U>0$, we do not find
a smooth $V_{LR}\rightarrow 0$ limit.  Perhaps this is not so
unsurprising: we similarly 
have no expectation that the problem is perturbative
in the dot-lead coupling, $V_{dl}$.

\vskip .05in

\noindent{\bf Analysis of model:}
We first examine the particular case of $V_{dR}=V_{dL}$.  We
will later show how this can be generalized.  To argue that
the model is thus integrable we
recast the two lead Anderson model into an even/odd basis
via the transformation,
$$
c_{e/o} = (c_L \pm c_R)/\sqrt{2}.
$$
With this transformation the Hamiltonian becomes,
\begin{eqnarray}\label{eii}
{\cal H} &=& H_e + H_o \cr
{\cal H}_e &=& -i \sum_{\sigma}
\int^\infty_{-\infty} dx c^\dagger_{e\sigma} (x) \del_x c_{e\sigma} (x) 
+ \ep_d \sum_\sig n_\sig \cr
&& \hskip -.15in 
+ U n_\up n_\down + \Gamma^{1/2}( c^\dagger_e (0) d_\sig + {\rm h.c.})
+ V_{LR} c^\dagger_{e\sigma}c_{e\sigma}|_{x=0};\cr
{\cal H}_o &=& 
-i \sum_{\sigma}\int^\infty_{-\infty} dx c^\dagger_{o\sigma} (x) 
\del_x c_{o\sigma} (x) 
- V_{LR} c^\dagger_{o\sigma}c_{o\sigma}|_{x=0},
\end{eqnarray}
with $\Gamma = (V_{dL}^2+V_{dR}^2)^{1/2}$ the total dot-lead coupling.
Note that the dot only couples to the even electrons.

In changing to an even/odd basis we are still able to compute scattering
amplitudes of electronic excitations off the dot.  The even-odd excitations
we employ scatter off the dot with a pure phase, $\delta_e/\delta_o$.
The corresponding reflection (R)/transmission (T) probabilities of an electronic
excitation in the original basis are given by 
\begin{eqnarray}\label{eiii}
T/R &=& |(e^{i\delta_e} \mp e^{i\delta_o})|^2/4.
\end{eqnarray}
Because of the simplicity of the odd sector, $\delta_o$ is energy
independent and given by $\delta_o = 2\tan^{-1}(V_{LR} )$.  
The zero temperature linear response conductance is given
in terms of T by
$$
G = T = {4V^2_{LR}\over (V^2_{LR}+1)^2}{(e+q)^2 \over e^2+1},
$$
where by identifying $e = \cot (\delta_e/2 + \tan^{-1}(V_{LR}))$
and $q = -\cot (2\tan^{-1}(V_{LR}))$ we have recast G in a
Fano-like form.  We
now turn to the non-trivial computation of $\delta_e$.

To compute $\delta_e$ we argue $H_e$ is solvable via Bethe ansatz.
To do so we proceed as in \cite{wie} for the ordinary Anderson model.
As a first step we identify an appropriate basis of single particle
excitations with momenta $\{ k_j \}$.  These single particle
eigenstates scatter off the dot with a {\it bare} phase
$\delta (k) = -2\tan^{-1}(\Gamma (k-\epsilon_d)^{-1}+V_{LR})$.
We then proceed to
compute the scattering matrices of these excitations
via computing two particle eigenstates.  These scattering
matrices are identical to that of the ordinary Anderson model.
In particular they satisfy a Yang-Baxter relationship.  As such
multi-particle eigenstates can be constructed in a controlled fashion.
For a set of N particles, their momenta, $\{ k_j \}$, must satisfy the
following quantization conditions \cite{wie}:
\begin{eqnarray}\label{eiv}
e^{ik_j L + i \delta (k_j)} &=& 
\prod^M_{\alpha = 1} { g(k_j) - \lambda_\alpha + i/2 \over  
g(k_j) - \lambda_\alpha - i/2};\cr
\prod^N_{j = 1} {\lambda_\alpha - g(k_j) + i/2 \over  
\lambda_\alpha - g(k_j) - i/2} &=& - \prod^M_{\beta=1}
{\lambda_\alpha - \lambda_\beta + i \over 
\lambda_\alpha - \lambda_\beta - i},
\end{eqnarray}
where $g(k) = (k-\epsilon_d-U/2)^2/2U\Gamma $.
The $M$ auxiliary parameters, $\{ \lambda_\alpha \}$, arise in forming
a multiparticle state carrying total $S_z = (N-2M)/2$.  The 
integrability of ${\cal H}_e$, a new result, leads to a set of
quantization conditions identical to that of the original Anderson model
but for one difference: $\delta (k_j)$ has a different form.  This
difference however is determinative of the physics.

To compute the {\it dressed} scattering phase, $\delta_e$, of an
electronic excitation, we employ an argument used by Andrei in computing
the $T=0$ magnetoresistance arising from magnetic impurities in a bulk
metal \cite{andrei}.  The momentum, $p$, of an added electron
is determined by the quantization conditions of a periodic system
of size $L$ via $p=2\pi n/L$.  This momentum has two contributions,
one coming from the bulk of the system and one from the dot,
i.e.,
$$
p = 2\pi n /L = p_{\rm bulk} + p_{\rm impurity}/L.
$$
The contribution coming from the dot, necessarily scaling as $1/L$,
is to be identified with the scattering phase of the excitation off
the dot, i.e. $\delta = p_{\rm imp}$.

The components of the momenta, $p_{\rm bulk}/p_{\rm dot}$, are dressed
by the fact that the ground state of the dot-lead system is a filled
Fermi sea of N interacting electrons (interacting inasmuch as the dot $U$
is finite).  In the language of the Bethe ansatz, an N particle ground state
for $\epsilon_d > -U/2$ is formed from $N$ total $k's$, $N-2M$ of
them real, the remaining $2M$ complex.  The $2M$ complex $k$'s  are given
in terms of $M$ real $\lambda_\alpha$'s with each $\lambda$ specifying
two $k$'s via $k_\pm = x(\lambda ) \pm i y(\lambda )$,
with 
$x(\lambda )=U/2+\epsilon_d -\sqrt{U\Gamma}(\lambda +(\lambda^2+1/4)^{1/2})^{1/2}$,
$y(\lambda )=\sqrt{U\Gamma}(-\lambda +(\lambda^2+1/4)^{1/2})^{1/2}$.

Under the Bethe ansatz the $k$'s are not to be thought of as bare momenta
of electrons.  Rather the Bethe ansatz affects a spin-charge separation
with the $k$'s associated with charge excitations and the $\lambda$'s with
spin excitations.  To compute an electronic scattering phase we must
glue together contributions coming from the spin and charge sectors
\cite{long}.  In adding a spin $\up$ electron to the system we both add
a real $k$ excitation as well as a hole in the set of $\lambda$-excitations.
The electronic scattering phase is then given by
$$
\delta^\uparrow_e = p^\uparrow_{\rm imp} = p^{\rm charge}_{\rm imp} (k)
+ p^{\rm spin}_{\rm imp} (\lambda ).
$$
The method of computing impurity momenta is discussed in detail in
\cite{long}.  The impurity momenta are related in turn to the impurity
densities, $\rho_{\rm imp}(k)/\sigma_{\rm imp}(\lambda )$, of
the $k/\lambda $ excitations via
$$
\partial_k p^{\rm charge}_{\rm imp} = 2\pi\rho_{\rm imp};~~~~
\partial_\lambda p^{\rm spin}_{\rm imp} = -2\pi\sigma_{\rm imp}.
$$
$\rho_{\rm imp}$ and $\sigma_{\rm imp}$ are then governed by the
equations,
\begin{eqnarray}\label{ev}
\rho_{\rm imp} (k) &=& \Delta (k) + 
g'(k)\int^{\tilde{Q}}_Q d\lambda a_1(g(k)-\lambda)\sigma_{\rm imp}(\lambda );\cr
\sigma_{\rm imp} (\lambda ) &=& \tilde{\Delta}(\lambda ) - 
\int^{\tilde{Q}}_Q d\lambda' 
a_2(\lambda'-\lambda)\sigma_{\rm imp}(\lambda ')\cr
&&-\int^B_{-D}a_1(\lambda-g(k))\rho_{\rm imp}(k),
\end{eqnarray}
where $\tilde{\Delta}(\lambda ) = 
-\partial_\lambda \delta (x(\lambda )+iy(\lambda ))/\pi$ and
$a_n(x) = 2n/\pi (n^2+4x^2)$.
$Q/B$ mark the `Fermi surfaces' of the seas of $k$ and $\lambda$ excitations
while $\tilde{Q}$ is related to the band cutoff, $D$.  For the purposes
of this paper we are only interested in computing the scattering of
electrons at the Fermi surface.  At the Fermi surface, $\delta^\uparrow_e$
is given by
$$
\delta^\uparrow_e|_{\rm Fermi surface} = p^{\rm charge}_{\rm imp}(k=B)
+ p^{\rm spin}_{\rm imp}(\lambda = Q).
$$
The scattering of spin $\downarrow$ excitations can be handled via
a particle-hole transformation \cite{long}.

We point out that $\tilde\Delta (\lambda )$ does not have a smooth
$V_{LR} \rightarrow 0$ limit, a notable difference with the 
results found in Ref.\cite{bulka,hofstetter}.  We thus do not expect the problem
to perturbative in $V_{LR}$.  We also emphasize that $\rho_{\rm imp}$ and
$\sigma_{\rm imp}$ encode {\it all} degrees of freedom scaling as $1/L$
($L$ is the system size) including corrections to the conduction electron
density, and not merely those living on the dot.
We now compute the linear response conductance
at $T=0$.

\vskip .05in 
\noindent {\bf Linear response conductance at $H=0$}:
In \cite{gores}, two well developed Fano resonances as a function
of the gate voltage are reported.  The resonances, plotted in Figure
1, appear as asymmetric dips.  
To model these resonances, we need to take into account $V_{dR}\neq
V_{dL}$.  To implement the even/odd basis change
we then use $c_{e/o} = (V_{dL/R}c_L \pm V_{dR/L}c_R)/\sqrt{\Gamma}$.
To assure the even and odd sectors do not interact 
we must allow an additional term to appear in the Hamiltonian,
$\delta H = (V_{dL}^2-V_{dR}^2)(c^\dagger_Lc_L-c^\dagger_Rc_R)/\Gamma$.
This term produces lead-lead backscattering.  For weak asymmetries
between $V_{dL}$ and $V_{dR}$, it should not unduly affect the physics.

\begin{figure}[tbh]
\centerline{\psfig{figure=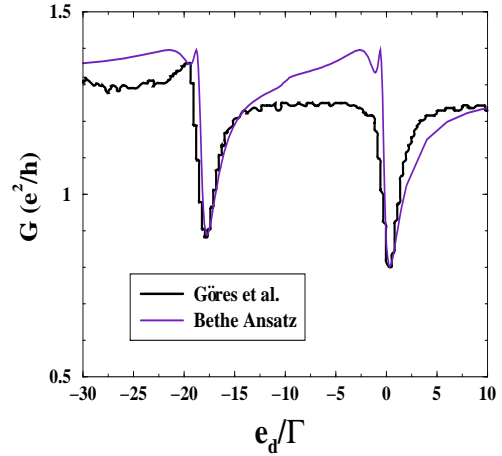,height=2.45in,width=2.5in}}
\caption{A plot describing the $T=0$ linear response conductance.
The solid
curve corresponds to that predicted by the Bethe ansatz given the 
parameters $\Gamma = .05 U$, $V_{Ld}/V_{Rd} = .75$, and $\Theta = 0.78$,
while the circles correspond to experimental data reported 
for a pair of Fano resonances reported in G\"ores et al. [2].}
\end{figure}

We now determine the necessary parameters, $U$, $\Gamma$, $V_{dL/R}$, and $V_{LR}$ 
entering the model Hamiltonian.
From the spacing of the
peaks together with their widths, the ratio $\Gamma /U$ is given by $1/20$.
To determine the values of $V_{Ld}/V_{Rd}$ and $V_{LR}$ we use
the fact that $G$ for large values of the gate voltage
tends to its $U=0$ value of $\gamma 4\Theta^2/(1+\Theta^2)^2$,
with $\gamma = 4V_{dL}^2V_{dR}^2/\Gamma^2$ and
$\Theta = V_{LR}/\sqrt{\gamma}$.
Furthermore the value of $V_{LR}$ determines the depth of the dip
in $G$.  Using then only {\it two} data points we fix these later
two parameters.

With these 
in hand, we manage to reproduce the linear
response conductance for the entire span of both
peaks (see Figure 1).  We note that the data to which 
we compare our theoretical
calculations was taken at a low but finite temperature.  We expect
the extremely sharp features seen in our $T=0$ computation
to be washed out at this temperature.

A feature of these resonances is that $G$ never vanishes.  
In terms of our computation we believe this 
to be the result of our non-perturbative treatment of
finite Coulomb interactions.  In the free case
$G(\epsilon_d )=\gamma \sin^2(\tan^{-1}(\Theta
- \Gamma/\epsilon_d)+\tan^{-1}(\Theta))$ and $G$ vanishes for
some $\epsilon_d$.  Similarly the expression
for $G$ arising out of the Dyson equations \cite{bulka,hofstetter}
always vanishes  for some value of $\epsilon_d$, one reason we suspect
that the Dyson equations do not adequately capture the physics at
finite $U$, $V_{LR}$.  We also point out that the resonances
occur for values of the gate voltage placing the dot in 
its mixed valence regime ($n_d < 1$) and not the Kondo regime ($n_d \sim 1$).
Generically, our solution predicts that the linear response conductance
in the Kondo regime will be relatively structureless.

\begin{figure}[tbh]
\centerline{\psfig{figure=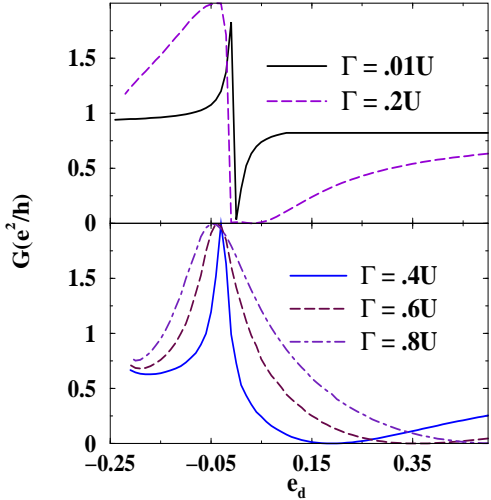,height=2.66in,width=2.5in}}
\caption{A set of Fano resonances for differing values of $\Gamma$.
These curves are computed using $V_{LR} = 0.4$.}
\end{figure}

\noindent {\bf Dependence of width of Fano resonances upon $\Gamma$}:
In \cite{gores}, Fano resonances were studied as a function of
the total dot-lead coupling strength, $\Gamma$, where it was observed
that the width of Fano resonances exhibit a non-monotonic dependence upon
$\Gamma$.  (In a dot with a single tunneling path, the width of a resonance
merely increases with $\Gamma$.)  Together with this non-monotonicity,
the overall shape and amount of asymmetry in the Fano resonances was observed
to be sensitive to the strength of $\Gamma$.

We can reproduce this array of behaviour.  Plotted in Figure 2 is the
linear response conductance for a set of differing $\Gamma$'s.  For
$\Gamma$ small, a Fano resonance appears as a sharply peaked bipolar structure.
As $\Gamma$ is increased, as na{\"\i}vely expected, the bipolar peak
broadens.  However at some critical value of $\Gamma \sim (.3-.4)U$,
the bipolar resonance is replaced by a narrow unipolar one.  
With further increases in $\Gamma$,
this resonance proceeds to broaden out.

\vskip .05in

\noindent {\bf Linear response conductance at $H\neq 0$}:
The behaviour of Fano resonances in magnetic fields was also studied
in \cite{gores}.  It was found that the resonances exhibited a marked
response to extremely small magnetic fields ($g\mu H/ \Gamma \sim 10^{-2}$).
In particular they demonstrated that upon application of $H$, a small bipolar
Fano resonance is transformed into a much larger unipolar structure
(see inset to Figure 3).  We are able to reproduce this phenomena 
(see main body of Figure 3).  For $H=0$, a small bipolar resonance
in $G$ is present.  With the introduction of a small field, a large
unipolar peak is superimposed over the bipolar structure.  As this
calculation is done at $T=0$, finite T should lead the two structures
to merge leaving a reasonable representation of the experimental data.

The strength of $H$ necessary 
to produce the unipolar peak is on the order of
a putative Kondo temperature, $T_k$, which at the symmetric point of
a single channel dot is estimated by, 
$T_k \sim \sqrt{U\Gamma}\exp (-U/8\Gamma )$.
This might suggest that in applying $H$, a resonance (or lack thereof)
due to the Kondo effect is destroyed.

\begin{figure}[tbh]
\centerline{\psfig{figure=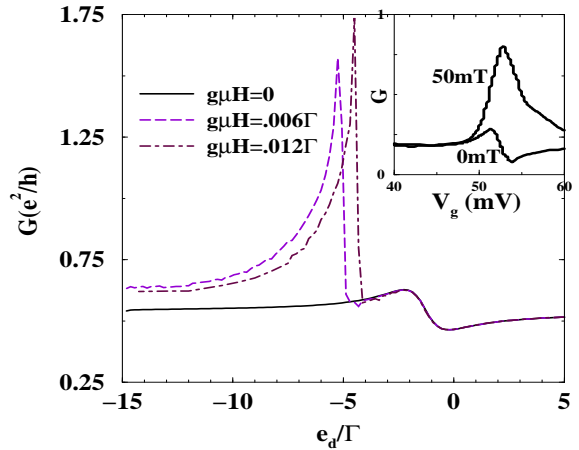,height=2.2in,width=2.9in}}
\vskip .15in
\caption{The response of a bipolar Fano resonance to the application
of small magnetic fields, $H$.  To compute these curves
we employ $\Gamma/U = 1/20$ and $V_{LR} = -3.6$.
Inset: Observed response
of a Fano resonance to small applied fields.}
\end{figure}

To summarize we have employed a generalized Anderson Hamiltonian to
describe observations of Fano resonances in quantum dots.  We have argued
that this Hamiltonian is integrable and sketched how this integrability
can be exploited to compute the $T=0$ linear response conductance.
Using this model, we are able to describe a number of observed features of
Fano resonances presented in \cite{gores}.  Our exact solution of the model 
suggests that the physics underlying the resonances is non-perturbative
in the presence of finite Coulomb repulsion on the dot.

The author acknowledges support 
from the NSF (DMR-9978074).  He also acknowledges useful discussions 
with Aashish Clerk and David Goldhaber-Gordon.


\begin{references}

\bibitem{gores}
J. G\"ores, D. Goldhaber-Gordon, S. Heemeyer, M. A. Kastner, H. Shtrikman,
D. Mahalu, and U. Meirav, Phys. Rev. B {\bf 62}, 2188 (2000).

\bibitem{zach} I. Zacharia, D. Goldhaber-Gordon, G. Granger, 
M. A. Kastner, Yu. Khavin,
H. Shtrikman, D. Mahalu, and U. Meirav, Phys. Rev. B {\bf 64}, 155311 (2001).

\bibitem{gold} D. Goldhaber-Gordon, et al.
cond-mat/9807233; D. Goldhaber-Gordon et al.,
Nature 391 (1998) 156.

\bibitem{kondo}
S. Cronenwett, T. Oosterkamp, and 
L. Kouwenhoven, cond-mat/9804211;
W.G. van der Wiel et al., Science 289, 2105 (2000);
D.C. Ralph, R.A. Buhrman, PRL 72, 3401 (1994).

\bibitem{fano} U. Fano, Phys. Rev. 124, 1866 (1961).

\bibitem{madhavan} V. Madhavan, W. Chen, T. Jamneala, M. F. Crommie,
and N. Wingreen, Science 280, 567 (1998).

\bibitem{ujsaghy} O. \'Ujs\'aghy, J. Kroha, L. Szunyogh, and
A. Zawadowski, PRL 85, 2557 (2000).

\bibitem{kobayashi} A. Yacoby, et al., PRL 74, 4047 (1995);
K. Kobayashi et al., cond-mat/0202006.

\bibitem{clerk} A. Clerk et al., PRL 86, 4636 (2001).

\bibitem{bulka} B. Bulka et al., PRL 86 (2001) 5128.

\bibitem{hofstetter} W. Hofstetter et al., PRL 87, 156803 (2001).

\bibitem{langreth} D. Langreth, Phys. Rev. 150, 516 (1966).

\bibitem{wie} P. Wiegmann et al., Sov. Phys. JETP Lett. 
35 (1982) 77; N. Kawakami and A. Okiji, Phys. Lett. A 86 (1982) 483.

\bibitem{long} R. M. Konik, H. Saleur, and A.W.W. Ludwig,
PRL 87, 236801 (2001); ibid, cond-mat/0103044, in press with PRB.

\bibitem{andrei} N. Andrei, Phys. Lett. A. 87 (1982) 299.

\end{references}
\end{document}